\documentclass[article,preprintnumbers,superscriptaddress,showpacs]{revtex4-1}

\usepackage{graphicx}
\usepackage{bm}
\usepackage{amsmath}
\usepackage{amssymb}
\usepackage{slashed}
\usepackage[colorlinks,pdfstartview=FitH]{hyperref}
\hypersetup{linkcolor=blue,citecolor=blue,filecolor=black,urlcolor=blue}

\def \be {\begin{equation} }
\def \ee {\end{equation}}

\def \bem {\begin{multline}}
\def \eem {\end{multline}}

\def \bes {\begin{subequations} }
\def \ees {\end{subequations}}

\def \pd {\partial}

\def \a {\alpha}
\def \b {\beta}
\def \c {\chi}
\def \d {\delta}
\def \e {\epsilon}
\def \ce {\varepsilon}
\def \g {\gamma}

\def \o {\omega}
\def \p {\pi}
\def \ps {\psi}
\def \ph {\phi}
\def \r {\rho}
\def \l {\lambda}
\def \m {\mu}
\def \n {\nu}
\def \s {\sigma}

\def \th {\theta}
\def \x {\xi}
\def \h {\eta}

\def \D {\Delta}
\def \G {\Gamma}

\def \Sg {\Sigma}
\def \tO {\tilde{\Omega}}
\def \tP {\tilde{P}}
\def \pr {\parallel}
\def \pp {\perp}

\def \<{\langle}
\def \>{\rangle}
\def \+{\dagger}
\def \({\left(}
\def \){\right)}
\def \[{\left[}
\def \]{\right]}

\def \tr {\text{tr}}

\def \Im {\text{Im}}

\usepackage[normalem]{ulem}

\begin{document}

\author{Defu~Hou} 
\affiliation{Institute of Particle Physics (IOPP) and Key Laboratory of Quark and Lepton Physics (MOE),  Central China Normal University, Wuhan 430079, China}
\author{Shu~Lin}
\affiliation{School of Physics and Astronomy, Sun Yat-Sen University, Zhuhai 519082, China}

\title{Polarization Rotation of Chiral Fermions in Vortical Fluid}
\date{\today}

\begin{abstract}
The rotation of polarization occurs for light interacting with chiral materials. It requires the light states with opposite chiralities interact differently with the materials. We demonstrate analogous rotation of polarization also exists for chiral fermions interacting with quantum electrodynamics plasma with vorticity using chiral kinetic theory. We find that the rotation of polarization is perpendicular both to vorticity and fermion momentum. The effect also exists for chiral fermions in quantum chromodynamics plasma with vorticity. It could lead to generation of a vector current when the probe fermions contain momentum anisotropy.

\end{abstract}
\maketitle 

\section{Introduction}%

It is known that polarized light interacting with stereoisomers can lead to rotation of polarization \cite{LANDAU1984358}. The polarization rotation effect has received much attention in different fields including optics \cite{PhysRevB.79.035407}, condensed matter physics \cite{PhysRevLett.116.077201}, cosmology \cite{Ni:2007ar} etc. The mechanism of rotation of polarization is that light with opposite circular polarizations interact differently with the chiral materials, leading to circular birefringence. The polarization dependent interaction is not particular for light. A natural question to ask is whether analogous effect exist for chiral fermions?

In this letter, we give one such example with chiral fermions interacting with polarized medium. Our medium is polarized by vorticity of fluid. On general ground, the spin polarization of chiral fermion in local rest frame of the fluid can be decomposed as follows
\begin{align}\label{decomp}
  {\cal P}^i=A_1\hat{p}^i+A_2\o^i+A_3\e^{ijk}\hat{p}_j\o_k,
\end{align}
with $\hat{p}^i$ and $\o^i$ being the direction of momentum and fluid vorticity. By parity, $A_1$ and $A_3$ are pseudoscalar functions and $A_2$ is a scalar function. In fact, $A_1$ exists in free theory due to spin-momentum locking for chiral fermion, and $A_2$ is a manifestation of spin-orbit coupling giving rise to chiral vortical effect \cite{Vilenkin:1980zv,Erdmenger:2008rm,Banerjee:2008th,Neiman:2010zi,Landsteiner:2011cp}. $A_3$ is a new contribution we will focus on.

The new contribution leads to net vector current in systems with momentum anisotropy. To see that, we note momentum integration of ${\cal P}^i$ gives rise to axial current, which receives opposite contribution from left and right-handed components of chiral fermion due to the odd parity of $A_3$. It follows immediately that a net vector current is generated as
\begin{align}\label{}
  J_V^i\sim \e^{ijk}\x_j\o_k,
\end{align}
with $\x_j$ being the axis characterizing the momentum anisotropy. This is analog of Lorentz force due to vorticity.



\section{Kinetic theory for chiral fermions}%

We illustrate this effect in weakly coupled quantum electrodynamics (QED) plasma using kinetic theory. We will generalize to quantum chromodynamics (QCD) plasma later. In high temperature limit of the two cases, electron/quark are approximately chiral. While the spin averaged kinetic theory has been widely used in describing transport coefficients of weakly coupled plasma \cite{Blaizot:2001nr,Arnold:2002zm,Arnold:2000dr,Arnold:2003zc}, its construction limits its application in spin dependent phenomenon, such as chiral magnetic effect \cite{Vilenkin:1980fu,Kharzeev:2004ey,Kharzeev:2007tn,Fukushima:2008xe} and chiral vortical effect \cite{Vilenkin:1980zv,Erdmenger:2008rm,Banerjee:2008th,Neiman:2010zi,Landsteiner:2011cp}. Spin dependent kinetic theory has been developed in recent years under the names of chiral kinetic theory \cite{Son:2012bg,Son:2012wh,Son:2012zy,Stephanov:2012ki,Pu:2010as,Chen:2012ca,Hidaka:2016yjf,Manuel:2013zaa,Manuel:2014dza,Huang:2018wdl,Liu:2018xip,Lin:2019ytz,Sheng:2017lfu,Lin:2019fqo} and spin kinetic theory \cite{Hattori:2019ahi,Wang:2019moi,Gao:2019znl,Yang:2020hri,Weickgenannt:2019dks,Gao:2019zhk,Liu:2020flb,Gao:2020vbh,Liu:2020ymh}, in which a scalar-like distribution function is used. In this paper, we retain the spinor structure and work with spinor equations. We start with the Kadanoff-Baym equation (KBE) \cite{Blaizot:2001nr}
\begin{align}\label{kbe}
&\frac{i}{2}\slashed{D} S^<(X,P)+\slashed{P}S^<(X,P)=
\frac{i}{2}\(\Sg^>(X,P)S^<(X,P)-\Sg^<(X,P)S^>(X,P)\),
\end{align}
where $\slashed{D}=\slashed{\pd}_X+ie\slashed{A}$. $S^{</>}$ are the Wigner transform of the off-equilibrium lesser and greater fermion correlators \footnote{$S^<(X,P)$ is related to the usual Wigner function $W(X,P)$ by $S^<=-\frac{1}{(2\p)^4}W$}
\begin{align}\label{Slg}
&S_{\a\b}^>(X,P)=\int d^4(x-y)e^{iP\cdot(x-y)}\<\ps_\a(x){\bar\ps}_\b(y)\>, \nonumber\\
&S_{\a\b}^<(X,P)=-\int d^4(x-y)e^{iP\cdot(x-y)}\<{\bar\ps}_\b(y)\ps_\a(x)\>,
\end{align}
with $X=\frac{x+y}{2}$ and similarly $\Sg^{</>}$ for lesser and greater self-energy correlators.

In the absence of collisional term on the right hand side (RHS), \eqref{kbe} is easily solved in a gradient expansion assuming $P\gg \pd_X$. Denoting zeroth order and first order solutions by $S^{<(0)}$ and $S^{<(1)}$, we obtain the following equation
\begin{align}\label{eqn_free}
\frac{i}{2}\slashed{\pd}S^{<(0)}+\slashed{P}S^{<(1)}=0.
\end{align}
The zeroth order solution $S^{<(0)}$ is given by propagator in local equilibrium with vortical fluid
\begin{align}\label{S0}
S^{<(0)}=-(2\p)\slashed{P}\d(P^2)\e(P\cdot u)f(P\cdot u),
\end{align}
where $\e(P\cdot u)$ is the sign function, $u$ is the fluid velocity and $f$ is Fermi-Dirac distribution function
\begin{align}
  f(P\cdot u)=\frac{1}{e^{(P\cdot u-\m)/T}+1},
\end{align}
with temperature $T$ and chemical potential $\m$ taken to be constants so that the plasma has vorticity only \footnote{If the system has net charge density, magnetic field can be induced by charge current. The effect of the induced magnetic field on the dynamics of fermions is suppressed by an additional power of $e^2$ compared to vortical effect considered here.}. With the vorticity counted as first order in gradient, \eqref{eqn_free} can be solved by \cite{Gao:2018jsi,Fang:2016vpj}
\begin{align}\label{S1}
S^{<(1)}=-(2\p)\frac{1}{2}\slashed{\tP}\g^5\d(P^2)\e(P\cdot u)f'(P\cdot u),
\end{align}
with $\tP_\m=P^\l\tO_{\l\m}$ and $\tO^{\m\n}=\frac{1}{2}\e^{\m\n\r\s}\pd_\r u_\s$. $\tO^{\m\n}$ can be decomposed into vorticity $\o^\m=\frac{1}{2}\e^{\m\n\r\s}u_\n\pd_\r u_\s$ and acceleration $\ce^\m=\frac{1}{2}u^\l\pd_\l u^\m$ as
\begin{align}\label{tO_rep}
\tO^{\m\n}=\o^\m u^\n-\o^\n u^\m+\e^{\m\n\r\s}\ce_\r u_\s.
\end{align}
We restrict ourselves to the case with only static vorticity in the local rest frame of the fluid. In this case, $\slashed{\pd}=\g^i\pd_i$ in \eqref{kbe}.
\eqref{S1} is the off-equilibrium correction to propagator due to fluid vorticity. The factor $\d(P^2)$ indicates that the on-shell condition is not changed.
We can infer the change of polarization due to vorticity. In local rest frame of the fluid, the unintegrated polarization is given by
\begin{align}
{\cal P}^i(X,\vec{p})=\int \frac{dp_0}{2\p}\tr \g^i\g^5 S^<.
\end{align}
\eqref{S0} corresponds to an unpolarized fluid. \eqref{S1} leads to a net polarization along the vorticity: ${\cal P}^i\sim f'(p)\o_i$ for both fermions and antifermions in neutral fluid.

Now we turn to the collisional term on the RHS. Recent works incorporating collisional term in spin-dependent theories include \cite{Yang:2020hri,Weickgenannt:2020aaf,Weickgenannt:2019dks,Sheng:2021kfc,Carignano:2019zsh,Li:2019qkf,Yamamoto:2020zrs,Wang:2020pej}. We use the following representation for the fermion self-energy \cite{Blaizot:1999xk}
\begin{align}\label{Sigma}
&\Sg^>(X,P)=e^2\int_Q\g^\m S^>(X,P+Q)\g^\n D_{\n\m}^<(X,Q) \nonumber\\
&\simeq -e^2\int_Q\g^\m S^>(P+Q)\g^\n D_{\n\a}^R(Q)
\Pi^{\a\b<}(Q)D_{\b\m}^A(Q),
\end{align}
with $\int_Q\equiv\int \frac{d^4Q}{(2\p)^4}$. We have suppressed the dependence on $X$ in $S^>$, $D^{R/A}$ and $\Pi^<$ in the last line for notational simplicity.
The representation is valid off-equilibrium, with the second equality holds to the leading order in gradient expansion, which requires $Q\gg \pd_X$.
A similar representation exists for $\Sg^<(X,P)$ with the exchange of $<$ and $>$ in \eqref{Sigma}. The off-equilibrium photon self-energy $\Pi^{\a\b}$ can be expressed in terms of fermion propagators as follows
\begin{align}\label{Pi}
\Pi^{\a\b<}(X,Q)=e^2\int_K tr\[\g^\a S^<(X,K+Q)\g^\b S^>(X,K)\].
\end{align}
In general, the KBE \eqref{kbe}, and the representations for self-energies \eqref{Sigma} and \eqref{Pi} do not form a closed set of equations as they also involve photon propagators, for which a separate kinetic theory for photons is needed. On the other hand, it is known that the RHS contains possible IR divergence \cite{Pisarski:1993rf,Blaizot:1996hd,Blaizot:1996az,Blaizot:1997kw}. If we keep only the leading IR divergence on the RHS, the kinetic theory for photons decouples for the following reason: we know the divergence comes from Coulomb scattering with soft photon exchange. The self-energy of soft photon $\Pi^{\a\b}$ as well as propagators $D_{\n\a}^R$ and $D_{\b\m}^A$ are entirely determined by hard fermion, which is governed by the kinetic theory. It will not be true if we wish to go beyond the leading IR divergence, for which Compton scattering is also needed \footnote{This can be seen from the following power counting. In Coulomb and Compton scatterings, the exchanged particles are photon and fermion respectively. The former contains $1/Q^4$ from two propagators of the exchanged photon while the latter contains only $1/Q^2$ from propagators of fermion. Moreover, the Coulomb scattering has an additional $1/Q$ due to the Bose enhancement from the photon self-energy.}. This allows us to use kinetic theory for fermion only to study the leading IR divergence. However, we will see that similar simplification is not possible in QCD where the presence of three-gluon vertex necessitates the inclusion of gluon kinetic theory for the same IR divergence.

It is known that correction to the $S^{>/<}$ and $\Sg^{>/<}$ corresponds to correction to hard fermion distribution, while correction to $D^{R/A}$ corresponds to correction to scattering amplitude \cite{Blaizot:1999xk}. We are interested in linear response to vorticity, thus the vortical correction can enter only one case. The vortical correction to $D^{R/A}$ is irrelevant because $S^{>/<}$ and $\Sg^{>/<}$ arises from equilibrium fermion distribution and the collision term vanishes identically by detailed balance independent of the interaction.

\section{Probe fermions in vortical fluid}%

Now we introduce probe fermions as a perturbation to the vortical fluid and study its spin polarization by solving the kinetic equation. We denote the perturbation to $S^<$ and $S^>$ by $\D S^<$ and $\D S^>$ respectively. In the quasi-particle approximation, we have $S^>(X,P)-S^<(X,P)=\r(X,P)=2\p\e(P\cdot u(X))\slashed{P}\d(P^2)$ \cite{Blaizot:2001nr}. The RHS is the local spectral density, which depends on local temperature and fluid velocity only, but not on the perturbation. It follows that $\D S^>(X,P)-\D S^<(X,P)=0$. Below we will use $\D S$ to denote both $\D S^<$ and $\D S^>$ and assume the on-shell condition is not changed, which will be verified by the explicit solution.

Now we work out the RHS of \eqref{kbe} up to first order in vorticity. 
At zeroth order, the RHS of \eqref{kbe} can be written as
\begin{align}\label{DS_combine}
&-\frac{i}{2}e^2\int_Q \big[\g^\m S^{>(0)}(P+Q)\g^\n D_{\n\m}^{<(0)}(Q)-\g^\m S^{<(0)}(P+Q)\g^\n D_{\n\m}^{>(0)}(Q)\big]\D S(P).
\end{align}
Note that the leading IR divergence comes from exchange of soft photon with $Q\ll P\sim T$. We can then approximate the equilibrium photon propagators as $D_{\n\m}^<(Q)\simeq D_{\n\m}^>(Q)=\frac{T}{Q\cdot u}(u_\m u_\n\r_L+P_{\m\n}^T\r_T)$ in Coulomb gauge with $u_\m u_\n$ and $P_{\m\n}^T$ being the longitudinal and transverse projection operators, and $\r_{L/T}$ being longitudinal and transverse spectral densities. We can then simplify the RHS as
\begin{align}\label{DQ_combine}
-\frac{i}{2}e^2\int_Q \frac{T}{Q \cdot u}\g^\m \r(P+Q)\g^\n D_{\n\m}^{<(0)}(Q)\D S(P),
\end{align}
Using $Q\ll P$ and $\D S\propto \d(P^2)$ and contracting the gamma matrices using
\begin{align}
\g^\m\g^\l\g^\n=g^{\m\l}\g^\n-g^{\m\n}\g^\l+g^{\m\l}\g^\n+i\e^{\m\l\n\s}\g_\s\g^5,
\end{align}
we obtain the RHS in local rest frame of the plasma as
\begin{align}
-\frac{i}{2}e^2\int_Q&\frac{T}{q_0}\(\frac{2p_0}{q^2}(Q^2\g^0-q_0\slashed{Q})\r_T+2p_0\g^0\r_L\)\d(2P\cdot Q)\D S(P)
\end{align}
We can perform the angular integration in the momentum integral to arrive at
\begin{align}\label{ang_int}
-\frac{i}{2}e^2\g^0\int\frac{dq_0qdq}{(2\p)^2}\frac{T}{q_0}\(\frac{q_0^2-q^2}{q^2}\r_T-\r_L\)\D S(P),
\end{align}
where we have used the on-shell condition $\d(P^2)$. 
The $q_0$ integral can be performed by using the sum rule \cite{ValleBasagoiti:2002ir}, but it is not necessary as we only need the leading divergence. Note that the longitudinal and transverse components correspond to electric and magnetic interactions respectively. The former is dynamically screened by the plasma to give finite contribution and the latter is only partially screened with divergent contribution. The leading divergence is from the kinematic regime $q_0\ll q$. We can approximate the retarded soft transverse correlator $\D_T$ and spectral density as \cite{Bellac:2011kqa}.
\begin{align}\label{D_T}
&\D_T(q_0\ll q)\simeq \frac{1}{q^2-i(\p q_0/4q)m_D^2},\nonumber\\
&\r_T\simeq 2\Im \D_T=\frac{1}{q^4+(\p q_0/4q)^2m_D^4}(\p q_0/2q)m_D^2.
\end{align}
Keeping only the leading divergence, we obtain from \eqref{ang_int}
\begin{align}\label{0th}
-\frac{i}{2}\g^0\frac{e^2T}{2\p}\ln\frac{m_D}{q_{\text{IR}}}\D S\equiv -i\g^0\G_0\D S.
\end{align}
Here $q_{\text{IR}}$ is an IR cutoff of momentum $q$. A resummation can be used to render the result finite \cite{Blaizot:1996hd}. We will not attempt it here as the IR regularized result is sufficient to illustrate the effect we are after.
Clearly the zeroth order contribution \eqref{0th} is independent of the spin as expected.

Now we turn to first order vortical correction to the collisional term, for which spin-dependent kinetic theory must be used. We first derive vortical correction to the collisional term \eqref{DS_combine}, which can enter either through $S^{>/<}$ or $D_{\n\m}^{</>}$. The former and the latter can be regarded as vortical correction to fermion and photon in the fermion self-energy loop respectively, or in language of kinetic theory, the former corresponds to the final state of the probe fermion and the latter corresponds to initial and final state of the medium fermion. We will show that the former vanishes identically and the latter give similar type of divergence as \eqref{0th}.

Let us work out the basic elements we need. We already have $S^{<(1)}$. We can solve for $S^{>(1)}$ from the following collisionless kinetic theory for $S^>$
\begin{align}
\frac{i}{2}\slashed{\pd}S^>+\slashed{P}S^>=0,
\end{align}
with the zeroth order solution
\begin{align}
  S^{>(0)}=-(2\p)\slashed{P}\d(P^2)\e(P\cdot u)(f(P\cdot u)-1).
\end{align}
Since $S^<$ and $S^>$ satisfy the same equation and the zeroth order solutions are related by the replacement $f\to f-1$, we easily obtain $S^{>(1)}=S^{<(1)}$ by analogy of \eqref{S1}. We now work out the vortical correction to the photon propagator $D_{\n\m}^{<(1)}$($D_{\n\m}^{>(1)}$). As we already show before, vortical correction to $D^R_{\n\a}$ and $D^A_{\b\m}$ are not relevant as they lead to vanishing collisional term, thus we only need to consider vortical correction to self-energy $\Pi^{\a\b<(1)}$, which is easily constructed using $S^<$ and $S^>$ as
\begin{align}\label{Pi1}
\Pi^{\a\b<(1)}(Q)=e^2\int_K tr\big[\g^\a S^{<(1)}(K+Q)\g^\b S^{>(0)}(K)+\g^\a S^{<(0)}(K+Q)\g^\b S^{>(1)}(K)\big].
\end{align}
Using \eqref{S0} and \eqref{S1}, we obtain
\begin{align}
&\Pi^{\a\b<(1)}(Q)\simeq2i(2\p)^2e^2\e^{\a\n\b\l}\int_K K^\m\tO_{\m\n}K_\l\d(K^2)\d(2K\cdot Q)f'(K\cdot u),
\end{align}
where we have used for soft photon momentum $Q\ll K$, $\e((K+Q)\cdot u)\simeq \e(K\cdot u)$ and $\d((K+Q)^2)\simeq\d(2K\cdot Q)$. Note that $\Pi^{\a\b<(1)}$ is antisymmetric in the indices. 
We consider $\Pi^{\a\b<(1)}$ in a fluid with vorticity but no acceleration, i.e. $\tO_{\l\r}=\o_\l u_\r-\o_\r u_\l$. Completing the momentum integral, we obtain in local rest frame of the fluid the following nonvanishing components for the vortical correction to the photon self-energy:
\begin{align}\label{Pi_exp}
\Pi^{ij<(1)}&=-\frac{ie^2}{4\p q}\c\(-\e^{ijk}\hat{q}_k\hat{q}\cdot\o\frac{q_0^2}{q^2}-\frac{1}{2}\e^{ijk}P_{kl}^T\o_l\frac{q^2-q_0^2}{q^2}+\e^{ijk}\o_k\),\nonumber\\
\Pi^{0i<(1)}&=-\frac{ie^2}{4\p q}\c\(-\e^{ijk}\hat{q}_j\o_k\frac{q_0}{q}\),
\end{align}
with $\c=-\int dk k^2\sum_{k_0=\pm k}f'(k_0)=\frac{\p^2T^2}{3}+\m^2$. 
Note that $\Pi^{\a\b<(1)}\sim O(1/q)$. This is reminiscent of Bose-enhancement in $\Pi^{\a\b<(0)}$.
The counterpart of $\Pi^{\a\b>(1)}$ can be obtained by the exchange of $>$ and $<$, which leads to $\Pi^{\a\b>(1)}=-\Pi^{\a\b<(1)}$, and thus $D_{\n\m}^{<(1)}=-D_{\n\m}^{>(1)}$.

The properties $D_{\n\m}^{<(1)}=-D_{\n\m}^{>(1)}$ and $S^{>(1)}=S^{<(1)}$ allow us to simplify the vortical correction to RHS of \eqref{kbe} as 
\begin{align}
&\frac{i}{2}e^2\int_Q\big[\g^\m S^{(1)}(P+Q)\g^\n \r_{\n\m}(Q)-\g^\m\(S^<(P+Q)+S^>(P+Q)\)\g^\n D_{\n\m}^{<(1)}\big]\D S(X,P).
\end{align}
The two terms correspond to vortical correction to final state of the probe fermion and to initial and final state of the medium fermion respectively. Let us first show the former contribution vanishes. To see that, we first perform the contraction of indices to obtain
\begin{align}\label{Qintegrand}
&\g^\m S^{(1)}(P+Q)\g^\n \r_{\n\m}(Q)=\d(2P\cdot Q)\big[-\slashed{\tP}\r_L+2\tP_0 \g^0\r_L-\frac{2\tP_0}{q^2}(Q^2\g^0-q_0\slashed{Q})\r_T-\frac{2\tP\cdot Q}{q^2}\(\slashed{Q}-q_0\g^0\)\r_T\big].
\end{align}
To proceed, we perform partial angular integration as
\begin{align}
  &\int_Q\d(2P\cdot Q)=\int\frac{dq_0q^2dqd\cos\th d\ph}{(2\p)^4}\d(2p_0q_0-2pq\cos\th)\nonumber\\
  &=\int\frac{dq_0dq d\ph}{(2\p)^4}\frac{q}{2p}.
\end{align}
The terms proportional to $\r_L$ in \eqref{Qintegrand} by integration of $q_0$ because $\r_{L}(q_0)$ is an odd function. The term proportional to $\g^0$ in the first bracket of $\r_T$ vanishes for the same reason. For the second term in \eqref{Qintegrand}, we expand $\slashed{Q}$ as
\begin{align}\label{slashq}
\slashed{Q}=\g^0q_0-\g_\pr q_\pr-\vec{\g}_\pp\cdot \vec{q}_\pp=\g^0q_0-\g_\pr q_0p_0/p-\vec{\g}_\pp\cdot {\vec q}_\pp.
\end{align}
Combining with the $q_0$ outside, we find the integrand is either odd in $q_0$ or in $\vec{q}_\pp$, which vanishes upon integration of $q_0$ or $\ph$. The remaining second bracket vanishes for similar reasons. Therefore vortical correction to final state of the probe fermion vanishes. The vanishing of this contribution is a consequence of the special kinematics $Q\ll P$, in which case the momenta of both probe fermion and medium fermion are almost unchanged in Coulomb scattering. It follows that in the absence of vortical correction, the spin angular momentum is unchanged for probe and medium fermions. The vortical correction to the final state of probe fermion modifies the spin, thus is not allowed by angular mmentum conservation.


Now we move on to vortical correction to medium fermions, which is not forbidden by the angular momentum conservation because vortical correction can enter both the initial and final states of medium fermions. Using \eqref{Pi_exp} and the following representation for $D^R_{\m\n}$ and $D^A_{\m\n}$ in Coulomb gauge:
\begin{align}
D^R_{\m\n}=u_\m u_\n\D_L+P^T_{\m\n}\D_T,\quad D^A_{\m\n}=D^R_{\m\n}{}^*,
\end{align}
we obtain the following components of photon correlator relevant in our case
\begin{align}\label{combinations}
&P_{ik}^T\Pi^{kl<(1)}P_{lj}^T=\frac{-ie^2}{4\p q}\c\big[\(1-\frac{q_0^2}{q^2}\)P_{im}^TP_{jn}^T\e^{mnk}\o_k\big], \nonumber\\
&\Pi^{0l<(1)}P_{li}^T=\frac{-ie^2}{4\p q}\c\(-\e^{ijk}\hat{q}_j\o_k\frac{q_0}{q}\).
\end{align}
After completing angular integration in a similar way as above, we obtain
\begin{align}\label{medium}
&\frac{i}{2}e^2\int_Q\big[-\g^\m \(S^<(P+Q)+S^>(P+Q)\)\g^\n D_{\n\m}^{<(1)}\big]\D S(X,P) \nonumber\\
&\simeq\frac{ie^4}{4p}\e(p_0)(2f(p_0)-1)\c\int\frac{dq_0dq}{(2\p)^3}\big[\(\g^i\g^5p_0-\g^0\g^5p_i\) \frac{q_0^2-q^2}{q^2}\(\frac{q_0^2}{q^2}\o_i^{\pr}+\frac{q^2-q_0^2}{2q^2}\o_i^\pp\)|\D_T|^2 \nonumber\\
&+\g^i\g^5\frac{q_0^2}{q^2}\(p\o_i-p_i\o_\pr\)\(\D_T\D_L{}^*+\D_L\D_T{}^*\)\big].
\end{align}
We are only interested in the leading divergence from the magnetic interaction, i.e. the $|\D_T|^2$ term. Further noting the divergence comes from the regime $q_0\ll q$, we can use \eqref{D_T} and $\vec{\o}^\pp\cdot\vec{p}=0$ to simplify \eqref{medium} as
\begin{align}
&\frac{ie^4}{8p}\e(p_0)(2f(p_0)-1)\c\int\frac{dq_0dq}{(2\p)^3}\g^i\g^5p_0\o_i^\pp|\D_T|^2\D S \nonumber\\
&=\frac{ie^4}{8p}\e(p_0)(2f(p_0)-1)\frac{1}{(2\p)^3}\frac{4\c}{m_D^2}\ln\frac{m_D}{q_{\text{IR}}}\g^i\g^5p_0\o_i^\pp \D S\nonumber\\
&\equiv-i\G_1\g^k\g^5p_0\o_k^\pp\D S.
\end{align}
We have used the same IR cutoff $q_{\text{IR}}$ as in \eqref{0th}. Using $m_D^2=\frac{e^2}{\p^2}\c$ for QED, we obtain
\begin{align}
  \G_1=-\frac{e^2}{8p}\e(p_0)(2f(p_0)-1)\frac{1}{2\p}\ln\frac{m_D}{q_{\text{IR}}}.
\end{align}
Note that at $\m=0$, $\G_1>0$ and is invariant under $p_0\to-p_0$. This is consistent with the charge conjugation symmetry in neutral plasma.

Now we are ready to solve the full kinetic equation with the RHS given by the sum of \eqref{0th} and \eqref{medium}
\begin{align}\label{full_kbe}
\frac{i}{2}\g^0\pd_t\D S+\frac{i}{2}\g^i\pd_i\D S+\slashed{P}\D S=-i\g^0\G_0\D S-i\G_1\o_i^\pp\g^i\g^5p_0\D S.
\end{align}
We have splitted $\slashed{\pd}$ into temporal and spatial parts, with vortical correction to the LHS entering only through the spatial parts.
In the absence of vortical correction, \eqref{full_kbe} reduces to
\begin{align}
  \frac{i}{2}\g^0\pd_t\D S+\slashed{P}\D S=-i\g^0\G_0\D S.
\end{align}
It adopts the following solution
\begin{align}\label{sol_0th}
\D S_0=e^{-2\G_0t}(-2\p)\(f_V\slashed{P}+f_A\g^5\slashed{P}\)\e(p_0)\d(P^2).
\end{align}
Here $f_V$ and $f_A$ are distribution function for vector and axial components of the probe fermions, to be distinguished from distribution function $f$ for the medium fermions. In fact, $f_V$ and $f_A$ are allowed to be aribitrary functions of momenta at this stage.
The exponential factor $e^{-2\G_0t}$ indicates damping of probe fermion due to collision with medium fermions.
The vortical correction to $\D S$ can be splitted into two parts $\D S_1$ and $\D S_2$, which satisfy respectively
\begin{align}\label{eqn_1st}
&\frac{i}{2}\g^0\pd_t\D S_1+\frac{i}{2}\g^i\pd_i\D S_0+\slashed{P}\D S_1=-i\g^0\G_0\D S_1, \nonumber\\
&\frac{i}{2}\g^0\pd_t\D S_2+\slashed{P}\D S_2=-i\g^0\G_0\D S_2-i\G_1\o_i^\pp\g^i\g^5p_0\D S_0.
\end{align}
$\D S_1$ and $\D S_2$ arise due to spin-vorticity coupling and vortical correction to collision term respectively. 
The solution to the first equation of \eqref{eqn_1st} is a simple generalization of \eqref{S1}
\begin{align}
\D S_1=e^{-2\G_0t}(-2\p)\(\frac{1}{2}f_V'\slashed{\tP}\g^5-\frac{1}{2}f_A'\slashed{\tP}\)\e(p_0)\d(P^2).
\end{align}
Note that the above solution arise from spatial derivative on $\D S_0$. Its validity relies on $f_V$ and $f_A$ being functions of $P\cdot u$, i.e. in local equilbirum.
The second equation of \eqref{eqn_1st} adopts the following solution
\begin{align}\label{S2}
  \D S_2=e^{-2\G_0t}(-2\p)\(\slashed{V}+\g^5\slashed{A}\)\e(p_0)\d(P^2),
\end{align}
with
\begin{align}
V^\m=-\G_1\e^{\m\n\r\s}\o_\n^\pp P_\r u_\s f_V,\;\;A^\m=-\G_1\e^{\m\n\r\s}\o_\n^\pp P_\r u_\s f_A.
\end{align}
Since the equation involves no spatial derivative on $\D S_0$, the validity of \eqref{S2} only requires medium fermions to be in local equilibrium.
The full solution is given by
\begin{align}\label{full_sol}
\D S=e^{-2\G_0t}(-2\p)\e(p_0)\d(P^2)
  \big[&f_V\slashed{P}-f_A'\slashed{\tP}-f_V\G_1\e_{\m\n\r\s}\g^\m\o_\perp^\n P^\r u^\s\nonumber\\
    +&f_A\g^5\slashed{P}-f_V'\g^5\slashed{\tP}-f_A\G_1\e_{\m\n\r\s}\g^5\g^\m\o_\perp^\n P^\r u^\s\big].
\end{align}
The corresponding spin polarization is given by
\begin{align}\label{full_P}
&{\cal P}^i=e^{-2\G_0t}\sum_{p_0=\pm p}
\big[f_A\frac{2\e(p_0)p_i}{p}-f_V'\o_i+f_A\G_1\frac{2\e(p_0)\e^{ijk}\o_jp_k}{p} \big].
\end{align}
\eqref{full_P} has the expected structure in \eqref{decomp}. Note that $f_V$ and $f_A$ are scalar and pseudoscalar functions respectively, which is consistent with the parity of $A_1$, $A_2$ and $A_3$ in \eqref{decomp}. The pseudoscalar function $f_A$ implies that the new contribution is opposite for left and right-handed fermions.

\section{Generalization to QCD plasma}


The polarization rotation effect can be generalized to QCD plasma by similar derivations. The new ingredients to take into account are the color factors and scattering processes not present in QED. The color structure modifies the KB equation only marginally: both the quark and gluon propagators are diagonal in color space:
\begin{align}\label{color}
  S^<_{IJ}\propto \d_{IJ},\quad D_{\m\n}^{ab<}\propto \d_{ab},
\end{align}
with $I,J$ and $a,b$ being indices in fundamental and adjoint representations of color group. The color sum in the quark self-energy reads $\sum_a t_{IJ}^at_{KI}^a=C_F\d_{JK}=\frac{N_c^2-1}{2N_c}\d_{JK}$. This amounts to the replacement $e^2\to g^2C_F$ in \eqref{Sigma} for quark self-energy.

The gluon self-energy contains both quark loop and gluon loop contributions. The former involves a replacement $e^2\to g^2N_f/2$ in \eqref{Pi} for $N_f$ flavor of quarks. The gluon loop contribution has no analog in QED. For the self-energy $\Pi^{</>}$, the four-gluon vertex is excluded. A further simplification can be made by noting that in Coulomb gauge contribution from unphysical polarizations and ghosts cancel each other in self-energy in equilirium. We assume it is still true for the off-equilibrium self-energy in the vortical plasma, hence we keep only transverse polarizations of gluons in the loop.

To proceed, we need the off-equilbrium propagator of gluon in vortical plasma. This is provided by solution of KB equation for gluons in Coulomb gauge \cite{Blaizot:2001nr}
\begin{align}\label{KB_gluon}
  \big[\(-i\pd\cdot P-P^2\)g^{\m\n}-P^\m P^\n+\frac{i}{2}\(\pd^\m P^\n+\pd^\n P^\m\)\big]D^{ab<}_{\n\r}=0,
\end{align}
with the Coulomb gauge condition
\begin{align}\label{gauge_cond}
  P^{\m\a}\(\frac{1}{2}\pd_\a-iP_\a\)D^{ab<}_{\m\n}=0.
\end{align}
Here $P^{\m\n}=-g^{\m\n}+u^\m u^\n$. The zeroth order solution to \eqref{KB_gluon} is simply the transversely polarized gluon propagator
\begin{align}\label{D0}
  D^{ab<(0)}_{\n\r}(X,P)=2\p P^T_{\n\r}\d(P^2)\e(P\cdot u)f_b(P\cdot u)\d^{ab},
\end{align}
with $f_b$ being the Bose-Einstein distribution function. The first order correction is given by \cite{Huang:2020kik,Hattori:2020gqh}
\begin{align}\label{D1}
  D^{ab<(1)}_{\m\r}(X,P)=-i(2\p)P_{\m\l}P_{\r\h}\frac{P^\l\pd^\h-P^\h\pd^\l}{2(P\cdot u)^2}\e(P\cdot u)f_b(P\cdot u)\d(P^2)\d^{ab}.
\end{align}
They give the following correction to the gluon self-energy through the gluon loop
\begin{align}\label{gluon_loop}
  &\Pi^{ab,\a\b<(1)}(Q)=-\frac{1}{2}g^2\int_K\(D_{\m\r}^{ce<(1)}(K+Q)D^{df>(0)}_{\s\n}(K)+D_{\m\r}^{ce<(0)}(K+Q)D^{df>(1)}_{\s\n}(K)\) \nonumber\\
  &\times f^{acd}[g^{\m\a}(-K-2Q)^\n+g^{\a\n}(Q-K)^\m+g^{\n\m}(2K+Q)^\a]f^{bef}[g^{\r\b}(K+2Q)^\s+g^{\b\s}(K-Q)^\r+g^{\s\r}(-2K-Q)^\b].
\end{align}
Let us focus on the contribution from the term proportional to $D_{\m\r}^{ce<(1)}D^{df>(0)}_{\s\n}$ for the moment. We can use $D^{ce>(0)}_{\s\n}\propto P^T_{\s\n}$ and transverse property of $P_{\s\n}^T$ to simplify the vertices. As before we only keep $Q$ in $\d((K+Q)^2)\simeq\d(2K\cdot Q)$ to arrive at
\begin{align}\label{loop1}
  \Pi^{ab,\a\b<(1)}_{10}(Q)
  &=-\frac{1}{2}g^2\int_KD_{\m\r}^{ce<(1)}(K+Q)D^{df>(0)}_{\s\n}(K)f^{acd}[-g^{\a\n}K^\m+2g^{\n\m}K^\a]f^{bef}[g^{\b\s}K^\r-2g^{\s\r}K^\b].
\end{align}
The subscript $10$ indicates it from the term $D_{\m\r}^{ce<(1)}D^{df>(0)}_{\s\n}$. The color factor is evaluated using $f^{acd}f^{bef}\d^{ce}\d^{df}=N_c\d^{ab}$.
The contraction of Lorentz indices reads
\begin{align}
  &P_{\s\n}^TP_{\m\l}P_{\r\h}[-g^{\a\n}K^\m+2g^{\n\m}K^\a][g^{\b\s}(K)^\r-2g^{\s\r}K^\b] \nonumber\\
  =&-P^{T,\a\b}\bar{K}_\l\bar{K}_\h-4P^T_{\l\h}K^\a K^\b+2P^{T,\b}_\l K^\a\bar{K}_\h+2P^{T,\a}_\h K^\b\bar{K}_\l,
\end{align}
with $\bar{K}_\m=-P_{\m\n}K^\n$. The first two terms vanish upon contracting with $K^\l\pd^\h-K^\h\pd^\l$ in $D_{\m\r}^{ce<(1)}(K+Q)\simeq D_{\m\r}^{ce<(1)}(K)$. The last two terms give
\begin{align}\label{Pi10}
  &\Pi^{ab,\a\b<(1)}_{10}(Q)\nonumber\\
  &=-\frac{i(2\p)^2}{2}g^2N_c\d_{ab}\int_K[-2K_T^\b K^\a\bar{K}\cdot\pd f_b(K\cdot u)+2\bar{K}^2K^\b\pd_T^\a f_b(K\cdot u)-(\a\leftrightarrow \b)]\frac{1+f_b(K\cdot u)}{2(K\cdot u)^2}\d(K^2)\d(2K\cdot Q) \nonumber\\
  &=-\frac{i(2\p)^2}{2}g^2N_c\d_{ab}\int_K[2\bar{K}^2K^\b\pd_T^\a f_b(K\cdot u)-(\a\leftrightarrow \b)]\frac{1+f_b(K\cdot u)}{2(K\cdot u)^2}\d(K^2)\d(2K\cdot Q),
\end{align}
where we have defined $K_T^\m=-P_T^{\m\n}\bar{K}_\n$ and $\pd_T^\m=-P_T^{\m\n}\pd_\n$. In the second line we have used $\bar{K}\cdot\pd f_b\propto \bar{K}^\m K^\n\pd_\m u_\n=0$ for fluid with vorticity only.

The evaluation of the term proportional to $D_{\m\r}^{ce<(0)}D^{df>(1)}_{\s\n}$ is similar by noting that $D^{df>(1)}_{\s\n}=D^{df<(1)}_{\s\n}$. It gives a contribution identical to \eqref{Pi10} except for the replacement $1+f_b(K\cdot u)\to -f_b(K\cdot u)$. Hence we obtain for their sum
\begin{align}\label{Pi1001}
  \Pi^{ab,\a\b<(1)}(Q)&=-\frac{i(2\p)^2}{2}g^2N_c\d_{ab}\int_K[2\bar{K}^2K^\b\pd_T^\a f_b(K\cdot u)-(\a\leftrightarrow \b)]\frac{1}{2(K\cdot u)^2}\d(K^2)\d(2K\cdot Q).
\end{align}
The vortical correction to the gluon self-energy from gluon loop is manifestly anti-symmetric in Lorentz indices as in quark loop counterpart. The explicit components are given by
\begin{align}\label{Pi1001_exp}
  \Pi^{ab,0i<(1)}&=-\frac{i(2\p)^2g^2N_c\d_{ab}}{2}\int_K k^2\frac{q_0}{q}P_T^{im}\e^{mnk}\hat{q}_n\o_k f_b'(k_0)\d(K^2)\d(2K\cdot Q), \nonumber\\
  \Pi^{ab,ij<(1)}&=\frac{-i(2\p)^2g^2N_c\d_{ab}}{2}\int_K\big[k^2\hat{q}_jP_T^{im}\e^{mnk}\hat{q}_n\o_k\frac{q_0^2}{q^2}+P_T^{im}P_T^{jn}\e^{mnk}\o_k\frac{k^2}{2}\(1-\frac{q_0^2}{q^2}\)\big]f_b'(k_0)-(i\leftrightarrow j).
\end{align}
Similar to the fermion loop case, only $\Pi^{ab,ij<(1)}$ enters the leading divergence we are after. Completing the momenta integral, we obtain the following combination
\begin{align}
  P^T_{im}P^T_{nj}\Pi^{ab,mn<(1)}
  =-\frac{ig^2N_c\d_{ab}}{8\p q}\c_0\(1-\frac{q_0^2}{q^2}\)P_T^{im}P_T^{jn}\e^{mnk}\o_k,
\end{align}
where $\c_0=\frac{\p^2T^2}{3}=\c(\m=0)$.
Comparing this with \eqref{combinations}, we find it is obtainable from QED case with $e^2\to g^2N_c/2$ and $\c\to\c_0$.

Now we are ready to combine the contribution from quark and gluon loops to give the following $\G_1$ for QCD
\begin{align}\label{G1}
  \G_1=-g^4C_F\frac{\e(p_0)(2f(p_0)-1)}{8p}\frac{1}{(2\p)^3}\frac{2(N_c\c_0+N_f\c)}{m_D^2}\ln\frac{m_D}{q_{\text{IR}}}.
\end{align}
For QCD, the Debye mass is given by
\begin{align}\label{debye}
  m_D^2=\frac{N_c\c_0+ N_f\c/2}{\p^2},
\end{align}
which allows us to express \eqref{G1} in a more suggestive form
\begin{align}\label{G2}
  \G_1=-g^2C_F\frac{\e(p_0)(2f(p_0)-1)}{8p}\frac{1}{4\p}\frac{N_c\c_0+N_f\c}{N_c\c_0+ N_f\c/2}\ln\frac{m_D}{q_{\text{IR}}}.
\end{align}
Note that the color and flavor factors do not cancel out. This has a simple explanation that the vorticity couples to quarks and gluons differently.

Let us apply the spin polarization \eqref{full_P} to early stage heavy ion collisions, where a significant vorticity is formed \cite{STAR:2017ckg}.
If the system is isotropic, \eqref{full_P} vanishes upon momentum integration. However for anisotropic system, which is also realizable at early stage of heavy ion collisions \cite{Kovchegov:2009he}, it could lead to an observable effect. As we emphasized above, $f_{V/A}$ can be arbitrary functions of momenta in \eqref{S2}. This could then lead to net current from the probe fermion with momentum anisotropy. For example, in a system with $f_A=0$ and anisotropic $f_V$, a vector current can be generated from the new contribution in $\D S$ as:
\begin{align}\label{current}
  \D J_V^i=\int_P\tr[\g^i\D S]\sim \e^{ijk}\x_j\o_k,
\end{align}
with $\x$ being the weighted average of momentum. A charge asymmetry for the probe quarks is needed such that the quark contribution to the current is not entirely canceled by the anti-quark counterpart.

\section{Conclusions and Outlook}%

We have found a new contribution to spin polarization due to interaction of chiral fermion with medium polarized by fluid vorticity. This is analogous to rotation of polarization in light interacting with polarized medium. We have considered the leading IR divergent part to the new contribution coming from Coulomb scattering between probe fermions and medium particles. The effect could arise due to vortical correction of final state of probe fermions or the counterpart of initial and final state of medium particles. We found the former contribution vanishes kinematically due to spin angular momentum conservation. For the latter contribution, the medium particles are fermions for QED plasma and are quarks and gluons for QCD plasma. The effect are qualitatively similar for the two cases. The fact that the vorticity couples differently to quarks and gluons is visible in a non-trivial color and flavor factor for the QCD case.

The effect we found is opposite for left and right-handed fermions, which cancels out in unpolarized probe fermions. Observation of this effect might be possible in polarized probe fermions. It is also desirable to generalize the current study to the case of massive fermions, where connection to spin polarization in heavy ion collisions can be made \cite{Becattini:2020ngo}. We expect the general structure in \eqref{decomp} remains unchanged.

The new contribution can also lead to net current in a system with momentum anisotropy. The current is perpendicular to both vorticity and axis characterizing the momentum anisotropy, which can be realized at early stage of heavy ion colisions. We defer more quantative analysis for future studies.

\begin{acknowledgments}
We are grateful to Jianhua Gao, Hai-cang Ren, Qun Wang, Di-Lun Yang, Ho-Ung Yee, Yi Yin and Pengfei Zhuang for useful discussions. This work is in part supported by NSFC under Grant Nos 11675274 and 11735007.
\end{acknowledgments}

\bibliography{vortical.bib}

\end{document}